# A framework to experiment optimizations for real-time and embedded software

H. Cassé[1], K. Heydemann[2], H. Ozaktas[2], J. Ponroy[3], C. Rochange[1], O. Zendra[3]

1: IRIT, Université de Toulouse, 118 route de Narbonne. 31062 Toulouse cedex 9, France
2: UPMC / LIP6, BC 167, Université Pierre et Marie Curie, 4 place Jussieu, 75252 Paris cedex, France
3: INRIA Nancy Grand Est - LORIA, 615 rue du jardin botanique, CS 20101, 54603 Villers-Lès-Nancy cedex, France

**Abstract**: Typical constraints on embedded systems include code size limits, upper bounds on energy consumption and hard or soft deadlines. To meet these requirements, it may be necessary to improve the software by applying various kinds of transformations like compiler optimizations, specific mapping of code and data in the available memories, code compression, etc. However, a transformation that aims at improving the software with respect to a given criterion might engender side effects on other criteria and these effects must be carefully analyzed. For this purpose, we have developed a common framework that makes it possible to experiment various code transformations and to evaluate their impact of various criteria. This work has been carried out within the French ANR MORE project.

## 1. Introduction

The design of general-purpose computing systems is often driven by performance targets and various hardware and software techniques can be used to meet requirements. Embedded systems differ from general-purpose systems by being subject to specific constraints: the code size may be limited by the capacity of the memory system; systems designed to be powered by batteries should exhibit low energy consumption; the inability to use efficient but voluminous cooling equipments may limit the allowed thermal dissipation. In addition, real-time embedded applications must meet hard or soft timing deadlines.

To meet these constraints, it might be necessary that the original software undergoes a series of transformations of all kinds. For example, some compiler optimizations might help to reduce the code size but more aggressive techniques like code compression may be helpful. The design space for transformations that can reduce energy requirements is very large, from compiler transformations that improve the efficiency and reduce the number of memory accesses to data placement strategies that aim at optimizing the use of the various memories available in the target hardware. The worst-case execution time – that must be analyzed to check that deadlines can be met for real-time tasks – can be reduced through different kinds of code transformations and mapping.

However, a transformation that aims at improving the code with respect to one criterion might sometimes impact another criterion due to possible side effects. These effects must be analyzed to check that all the constraints of the system can be met.

In an industrial context where time-to-market is important, being able to experiment several transformations in short time is desirable. This motivates the design of a dedicated framework which is one of the goals of the MORE (*Multicriteria Optimization for Real-time Embedded systems*)[1] project that started in 2007 and will end in 2010.

This paper describes the framework developed within this project and shows how it can be used to improve embedded software so that it meets its constraints as closely as possible. The presented framework is open to a large extent and has been designed to facilitate the implementation or emulation of new measurement tools as well as transformation tools. The tools implemented so far include an emulator for code compression, an interface to the gcc compiler to control some compiler optimizations used to improve the worst-case execution time, an emulator for data placement schemes, a cycle-level simulator, an energy estimator and a WCET estimator.

The paper is organized as follows. In Section 2, the overall architecture of the framework is described in details. Section 3 introduces the transformations we have considered so far to improve the code size, the energy consumption and the worst-case execution time of a piece of code. Section 4 is dedicated to evaluation (measurement and static

---

[1] The MORE project is supported by the French National Research Agency (ANR) under agreement n° ANR-06-ARFU-002

analysis) tools. In Section 5, we report the results of some experimental work carried out using our framework. Finally, concluding remarks and future work plans are given in Section 6.

## 2. Overview of the optimization framework

The goal of building a common software framework is to facilitate the experimentation of various kinds of code transformations so that their impact on different criteria can be analyzed and their respective results can be compared.

2.1 Supporting library

To make it possible to estimate and compare the impact of various code transformations on different criteria (execution time, energy consumption, etc.), it is desirable to have tools that work on a common representation of the code and of the target hardware architecture. This is why we have decided to build our framework on top of the OTAWA library that provides such facilities [4].

OTAWA is a library dedicated to the analysis of Worst-Case Execution Times for hard real-time tasks. It includes a number of tools to load a binary code (various target architectures are supported: PowerPC, ARM7, TriCore, Star12X), to decode it, and to build a representation of it in the form of a Control Flow Graph (CFG) where nodes stand for basic blocks and edges express the possible execution in sequence of two basic blocks. OTAWA also provides annotation facilities that allow hooking attributes to any code object (an instruction, a basic block, an edge between two blocks). Examples of user-defined annotations include timing information for a basic block or the behavior of an instruction with respect to the instruction cache. OTAWA is delivered under the LGPL license which makes it a chosen tool set for research work.

2.2 Implementation of code transformations

We distinguish two classes of transformations:

- some of them modify the structure of the CFG. For example, loop unrolling, function inlining, superblock construction add or concatenate some basic blocks.

- other ones keep the code structure but have an impact on the behavior of instructions. For example, code compression does not change the flow of instructions but modifies their fetch timing.

The first class mainly includes compiler transformations. One possibility to perform experiments with these transformations is to interface with the target compiler so that a new binary code including the effects of the transformation can be generated. Within the MORE project, we have successfully used the GCC-ICI interface [8] and developed specific plugins to control built-in compiler optimizations like loop unrolling and function inlining (see Section 3.3). Alternatively, the effects of transformations that modify the code structure could be emulated by rearranging the CFG representation within the framework. This would avoid re-building the code and would make the experiments faster.

The second class of transformations can be emulated by annotating the impacted instructions so that their new behavior or timing can be taken into account by measurement or analysis tools. For example, code compression can be expressed by annotating each instruction with the address it would have in the compressed code: this new address can be considered by a simulator or a code analyzer while the compressed binary code has not really been generated. In the same way, a transformation that would place data in specific memory can be emulated by annotating load/store instructions with their target memory. This way, there is no need to actually generate the transformed code to perform analyses. Using this approach, several kinds of transformations can be considered during the system design process, avoiding to really implement those that do not provide good results.

To summarize, our framework can host several kinds of facilities to emulate code transformations: CFG annotations, CFG manipulation and interfacing with the compiler.

2.3 Implementation of measurement and analysis tools

Our framework includes a cycle-level simulator built on top of SystemC. It can currently support targets with superscalar pipelines, in-order or out-of-order execution, branch prediction, instruction and data caches, user-specified memory architecture. If needed, the user can develop new components from generic modules to model specific hardware features. The simulator can be configured through an XML-like description of the hardware architecture. By default, the simulator provides the execution time of the code under analysis. However, it is possible to add software probes to get more specific measures that may be used to compute the code behavior with respect to various criteria. For example, statistics on the behavior of caches can be used to estimate the energy consumption of accesses to the memory hierarchy.

The framework also allows the implementation of static analysis algorithms through the concept of Code Processor provided in the OTAWA library. A code processor processes each node of a Control Flow Graph and uses annotations produced by previously executed code processors to generate, in turn, new annotations that improve the knowledge on the considered piece of code. Example code processors are pipeline and cache analyses used for the estimation of Worst-Case Execution Times.

In the next section, we will provide further details on the transformation and analysis tools that have been implemented within the MORE project.

## 3. Transformations

3.1 Energy-aware memory mapping

Energy-aware memory mapping consists in techniques and algorithms aiming at reducing the overall energy used in a computing system thanks to appropriate placement of information in the various kind of memories available in the system. The general idea is to take advantage of the different behaviors of these heterogeneous memories with respect to energy. These memories may be main memory (DRAM and its derivatives), cache memory (SRAM and its derivatives) or Scratch-Pad Memory (SPM, made of SRAM and its derivatives).

In the MORE project, we have so far focused our work pertaining to energy-aware memory mapping on the placement of data, not code, because the latter would have implied difficult to master interactions with code compression and impaired WCET analysis. We also focused our efforts on static placement issues. This means that the data is placed according to a layout decided before the actual execution takes place and can never change at runtime. Indeed, the context of real-time systems with WCET constraints makes it mandatory for us to be able to have a good predictability of timing, which dynamic algorithms tend to significantly impair.

The overall strategy to decrease energy consumption through memory placement roughly consists in trying to place the most accessed data in the less energy-hungry memories. We especially target SPM, since this kind of memory is very efficient energy-wise and very well-suited to embedded systems. Indeed, an SPM is basically a cache memory whose logic has been removed, leaving only the storage part. The management of the SPM is not performed in hardware, but by the executed program itself. Note that SPM are not to be mistaken with software caches. In a software cache, the cache logic is simply stripped out from the hardware to be put "as is" in software, which makes it very energy-hungry. In an SPM, the cache management algorithms do not exist anymore: the executed program explicitly takes care of the management of the SPM. This can be done manually by the program developer or with help from the compiler and runtime system.

In our work, we considered several transformations (or memory mapping strategies) to improve energy usage thanks to placement in SPM. SPM_firstUsed is a naive baseline approach and consists in placing data on a first-come, first served basis: the first accessed data in the program are placed in the SPM until it is full. This could be done on-the-fly, at low cost without prior knowledge of the program. SPM_smallSizeFirst consists in placing smaller data in SPM, then larger data if room remains. The idea is to maximize the number of data in SPM. Prior knowledge of all the data sizes is necessary to choose which data go to SPM. Finally, SPM_highFrequency consists in placing data in SPM by order of decreasing access frequencies. The idea is to maximize the number of accesses in SPM. Prior knowledge of all the access frequencies is necessary to choose the data that go to SPM.

To implement these transformations in the MORE project in a completely automated way, we extended the OTAWA simulator, as follows. First, a pre-run with the targeted program is performed, during which a trace of all accesses to each piece of data is recorded. This trace is then analyzed just before the actual run. For SPM_firstUsed, we extract the first accesses from the trace. For SPM_smallSizeFirst, the accesses are sorted by size. For SPM_highFrequency, the number of accesses to each piece of data is computed. This way, we can build a memory mapping for data according to the chosen strategy. Finally, when the program is actually run, this mapping is used by the simulator to re-route all memory accesses to data adequately. This makes it possible to smoothly compute the new WCET and energy usage as described in Section 4. Note that the data accesses we consider correspond to low-level, machine accesses, which are sub-structural with respect to the data structures in the source program.

3.2 Code compression

Code compression reduces the code size by compacting the original code into a non executable format [2]. At runtime, a decompression step is needed to retrieve the initial code. The previously proposed approaches differ in the compression strategy (statistical as Huffman coding, dictionary-based or any combination of both) as well as in the implementation (by software or in hardware) and in the location of the decompression engine [2]: between the cache and the memory for the pre-cache approaches [9][10], between the cache and the processor for post-cache schemes [1][13] or inside the processor core [5][16]. In the latter case, decompression is then very close to the translation engine for micro-coded instructions.

In the MORE project, we decided to use a post-cache or within-processor code compression technique that is likely to optimize at the same time the code size, the energy consumption and the performance contrary to pre-cache approach [13][16][10]. Indeed, as compressed code is stored in the instruction cache, it is likely to reduce the number of cache misses which might improve both the execution time and the energy consumption. Since our intention is to consider high-performance processors, we have opted for in-pipeline decompression since post-cache decompression is very hard to implement for superscalar processor and might impair the efficiency of a branch predictor. In addition, an in-pipeline approach avoids the complexity of handling different address spaces: the one related to the compressed code and the other one seen by the processor for which the code compression is completely transparent in case of pre-cache or post-cache decompression. Since the decompression overhead is critical because decompression may be needed at each cycle, we designed dictionary-based compression scheme that might be less efficient (in terms of compression rate) than statistical algorithms but that allows faster decompression.

In our solution, the dictionary contains full instructions. In order to limit the cost of the dictionary and to keep its access time short, it is desirable to restrict its size. Keeping the dictionary small is also necessary to limit the width of the dictionary index (log(n) bits are required for an n-entry dictionary), which is important to insure the efficiency of the code compression scheme: the smaller the index width, the better the compression rate. Moreover, a dictionary does not need to hold all the instructions that appear in the code: when an instruction in the dictionary appears only once in the code, the code size is not improved and even degraded (since the instruction is stored twice: once in the code, in a compressed form, and once in the dictionary).

As far as the dictionary does not hold all the instructions, the compressed code contains both compressed and uncompressed instructions. For our compression scheme design, we have fixed the dictionary size to 256 entries, which is a standard size for hardware implementation and one-cycle decompression [7]. Besides, this size allows covering a significant part of the static code and reaching good compression rate even with large applications (the most redundant instructions are generally not numerous) [13]. The main issue of a small dictionary-based compression scheme is how the dictionary is built. To maximize code size reduction, it is preferable to include the most statically repeated instructions of an application whereas selecting the most executed instructions favors the reduction of the number of instruction cache misses [13]. To trade-off the benefit from both code size and cache miss rate improvement, our compression scheme has one parameter P which controls over the dictionary building: P% of the dictionary is filled with the most executed instructions and the remaining entries are filled with the most statically repeated instructions.

Our compression scheme replaces two or three successive instructions present in the dictionary by one 32-bit encoding instruction, which is composed of an invalid operation code of the target ISA and indexes of the dictionary entries that store the corresponding instructions. Once the dictionary is built, sequences of two or three instructions that are in the dictionary and that belong to the same basic block (to avoid impairing branch prediction) are then selected to form an encoding instruction. Instead of producing a binary code, which is an error-prone (due to jump address patching) and so time-consuming process, instructions that are compressed are annotated as so. This information is sufficient to emulate decompression and to take into account compression in the measurement tools of our framework.

Decompression is done in the processor pipeline. A decompression stage must be added except if the processor already has a stage for translation of micro-coded instructions into instructions as in the Intel i686 architecture. The decompression stage is placed between the fetch and the decode stages. Non-compressed instructions are simply forwarded to the decode stage. In case of a compressed instruction, extra cycles are needed to access the dictionary. As the dictionary is much smaller and less complex than a cache, a one-cycle access is feasible. The dictionary access fills the pipeline with two or three new instructions depending on the number of instructions encoded into a single one.

3.3 Control of compiler optimizations

The transformations considered to improve the WCET estimates consist in making the code more linear (i.e. in removing flow control instructions) which improves the predictability of processor states. Such transformations are available in standard compilers: common examples are loop unrolling and function inlining optimizations.

The GCC Interactive Compilation Interface [8] has been designed to allow controlling the compiling process with limited intrusion in the compiler code: support is provided to develop plugins that can interact with GCC during the compilation process. In the MORE project, we have designed such plugins to control loop unrolling and function inlining: by the way of XML files, it is possible to specify which loop is to be unrolled, and by which factor, and which function is to be inlined. This makes it possible to apply the transformations only when they have a positive impact on the WCET estimates. It also allows curbing the increase of the code size in case of size-limited memory.

## 4. Evaluation tools

4.1 Energy consumption estimation

To carry our work, we extended the OTAWA functional simulator so that it provides energy consumption estimations. We decided to implement this in a simple hence robust way. We added counters to the simulator to retrieve read and write access numbers, for each memory component in our predefined architecture (be they cache memories, SPMs or DRAM memories). We also retrieve the unit costs in energy of read and write accesses, for each memory component. This way, with the access statistics and unit energy consumption, we are able to estimate the overall energy usage for our architecture as summarized in this formula:

$$E_{total} = \sum_{memories} N_{read\ access} * E_{read\ access} + N_{write\ access} * E_{write\ access}$$

We implemented the functional behavior of each memory component. For example in an architecture with DRAM and data cache, when the accessed data is not in data cache, then several accesses to DRAM occur to replace the corresponding cache line by the appropriate one, which increases the overall energy consumption. We are thus able to measure the impact of cache misses. New kinds of memory can thus easily be taken into account for energy, provided their functional behavior is added as well as the read and write counters.

The unit read and write energy consumption numbers are obtained by automatically calling an external tool called CACTI. First developed by Wilton and Jouppi, CACTI is an analytical model for the access and cycles times of on-chip direct-mapped and set-associative caches. CACTI takes many parameters (see Figure 1) into account when computing energy costs, like temperature, cache size, associativity, block size, transistor technology, etc.. It can also be used to compute the energy access costs for other kinds of memories like SPM or DRAM.

Even though read and write counters are enough to estimate energy consumption, they are not sufficient to understand and explain the results. Other counters were thus added. Cache hits, cache misses, cache read misses dirty, cache write misses dirty, SPM fails and successes make it possible to analyze and understand results evolution according to memory parameters or transformation parameters.

These numerous counters also allow us getting more detailed results, by showing the energy consumption not only globally but also for each memory component.

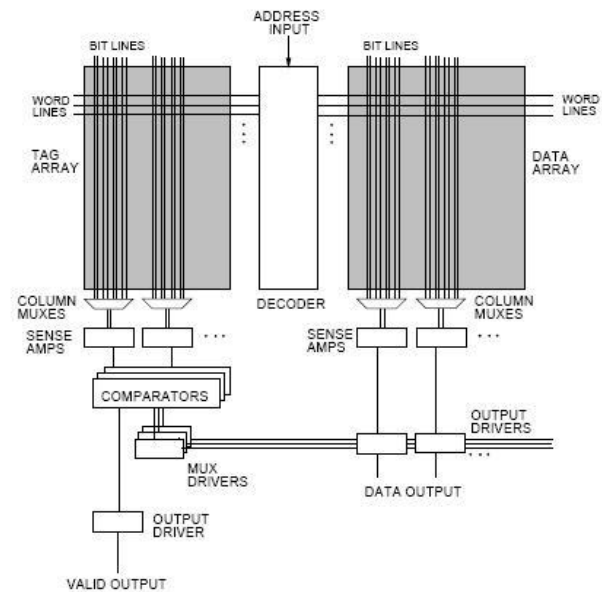

**Figure 1. Cache structure in CACTI (from [10])**

### 4.2 WCET analysis

To perform WCET analyses, we have developed a tool that uses several components available in the OTAWA library, including:

- a binary code loader and CFG builder
- an instruction cache analyzer based on abstract interpretation techniques [6][3]
- a data cache analyzer also based on abstract interpretation techniques
- a timing analyzer that evaluates the worst-case execution time of basic blocks taking into account the target architecture and the results of the instruction cache analysis [15]
- a flow-fact loader that reads flow fact annotations provided by the oRange tool [12]
- a WCET computer that builds an integer linear program according to the IPET method [11]. This program is solved using the lp_solve tool [18].

To estimate the impact of code compression on the WCET, we have extended the instruction cache analyzer so that it considers the instruction addresses in the compressed code (since the cache holds compressed code). Details about this extension can be found in [14].

To analyze the effects of data placement strategies, we have developed an interface to the data placement tool: through this interface, information about the memory in which each piece of data is stored (main memory, and then possibility in the data cache, or scratchpad memory) is transmitted to the data cache analyzer and WCET estimation tool.

## 5. Experimental validation

### 5.1 Methodology

To illustrate possible use of our framework, we have carried out some experiments with four test programs listed in Table 1. Two of them (adpcm and compress) belong to the benchmark collection dedicated to the estimation of WCET analysis tool maintained at the University of Mälardalen [19]. The others (helico and seg) have been developed during the project to fulfil our need to experiment on codes that exhibit timing analyzability and manipulate a sufficient amount of data. Their source code will be made available soon.

| | |
|---|---|
| adpcm | Adaptive pulse code modulation algorithm. |
| compress | Data compression program. |
| helico | Software that controls a toy helicopter for a mission including hovering. |
| segmentation | Image segmentation algorithm. |

**Table 1: Test programs**

We considered a system with a generic 2-way superscalar processor with in-order execution and a 2-way associative 1KByte instruction cache (we voluntarily selected a small cache size to get realistic results with our small test programs). For the data, we considered two configurations: the first one (Config1) includes a 1KByte data cache and the second one (Config2) a 512-Byte data cache and and a 512-Byte scratchpad memory. In both cases, the data cache is two-way set associative with LRU replacement policy.

### 5.2 Impact of a data placement strategy

In this section, we show and discuss the effects of the chosen data placement strategies on energy consumption, code size and worst case execution time. The 3 data placement strategies we considered, SPM_firstUsed, SPM_smallSizeFirst and SPM_highFrequency, were described in Section 3.1.

As could be expected, since SPM_firstUsed is a naïve baseline approach, our experiments showed that a Config2 hardware configuration with an SPM managed according to the SPM_firstUsed strategy increases the energy consumption for all benchmarks when compared to Config1. SPM_firstUsed is thus not a good strategy. The second strategy, SPM_smallSizeFirst, requires even more energy than the previous one. It is therefore an even worse transformation. These two strategies increase energy consumption because they are not tailored to the executed program but carved in stone, since they do not take into account the most frequently accessed data. Conversely, SPM_highFrequency is based on actual, observed accesses. This data placement decreases energy consumption from 8% to 66% on the four considered benchmarks (see Figure 2) when compared to Config1. Coupling a cache and an SPM is thus highly valuable with this strategy and performs betters than alone cache.

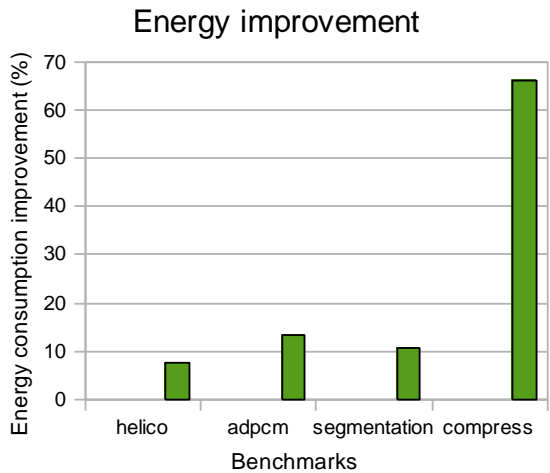

**Figure 2. Energy impact of SPM_highFrequency**

The memory placement strategies we consider are applied during the compilation stage, by modifying the memory mapping information that is used when a benchmark is loaded into memory for execution. Only the addresses are changed, not the binary itself, and more specifically not its instructions. In OTAWA, we emulate this behavior by keeping memory mapping and working only on data access addresses. Therefore, the impact of our data placement memory on code size is non existent. This give us all the liberty to choose the most appropriate memory transformation. Note that in fact, for convenience reasons, we do not actually change the addresses in the binary, but emulate this change in the OTAWA simulator.

Regarding to the effect on the WCET criterion, we only considered our SPM_highFrequency strategy, the other two being invaluable. As Table 2 shows, using a Config2 hardware configuration with an SPM managed by SPM_highFrequency improves the WCET for each considered benchmark. One reason for this improvement is because when considering an architecture with a data cache, the WCET computed by static analysis may be significantly overestimated, since it is not always possible to predict whether a piece of data is in the data cache or not when it is accessed. This inaccuracy comes from the facts that all possible ways are not explored and that static analysis techniques are used which work by state fusion at CFG junction points. Replacing (part of) a cache by an SPM (which has fixed latency) thus removes part of the uncertainty and makes it possible to have a less pessimistic, more accurate WCET. A second reason for this improvement of WCET is the fact that the overall execution time should be better because the number of cache hits plus number of SPM hits in Config2 is higher than the number of cache hits in Config1.

The segmentation benchmark is not considered here, because the WCET could be computed only on a subset of the program.

| Benchmark | Impact on WCET |
|---|---|
| adpcm | -23.8% |
| compress | -18.5% |
| helico | -5.9% |

**Table 2: WCET impact of SPM_highFrequency**

5.3 Impact of code compression

In this section, we show and discuss the effects of code compression on code size, energy consumption and worst-case execution time. As our compression scheme has a parameter fixing the percentage of the dictionary that is filled with the most executed instructions, we have tested its effects on criteria by varying its value. Both graphs in Figure 3 show the measured criteria for different values of P for two out of the four considered applications, namely segmentation and compress. These figures show that for compress, the higher the value of P, the better the ratio for energy consumption, WCET and ACET.

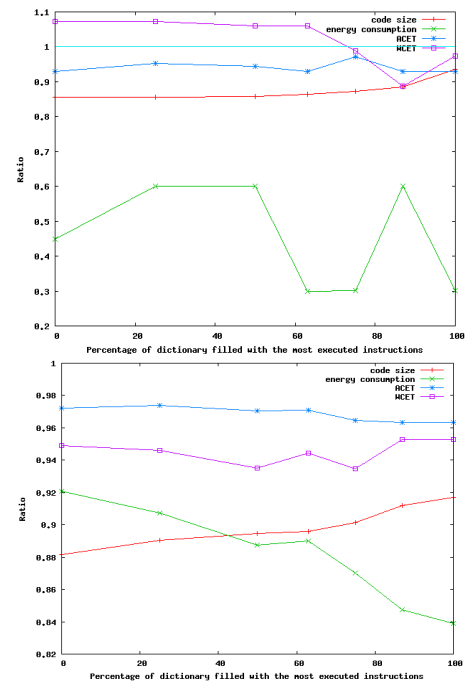

**Figure 3. Impact of the percentage of the dictionary filled with the most executed instruction on code size, energy, execution time (ACET) and WCET for segmentation (leftmost figure) and compress (rightmost figure).**

For all benchmarks, code size reduction is at its maximum when P is set to 0 and decreases as P increases. This is logical since a smaller P favors code size reduction whereas a greater P favors optimizing energy consumption and ACET. The WCET is always reduced but the improvement is not correlated to the P value.

However, for some benchmarks as shown for the segmentation application in Figure 3, there is no correlation between P and the effects of compression on energy consumption and WCET. The energy consumption can be worse with an higher value of P even if it should favor the compression of the most executed instructions and so should reduce the number of instruction cache misses. This is due to the fact that code compression changes the code placement which may increase the number of cache accesses depending on the alignment of instructions on cache line boundaries or increase conflict misses in the instruction cache. The WCET can be either improved or degraded by compression depending on the value of P.

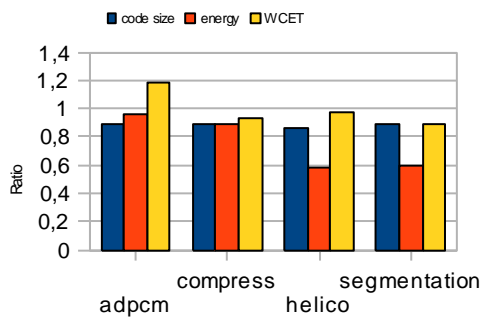

**Figure 4. Effect of code compression on code size, energy consumption and WCET**

For each application, we have chosen a value of P that either leads to a good trade-off between the three criteria or to the shortest WCET. The results are illustrated in Figure 4. For the selected values of P, energy consumption is reduced from 5% to 40%, code size from 11 to 13%. WCET is reduced by 6% for compress, by 11% for segmentation and is almost unchanged for helico (0.2% improvement). The WCET of adpcm degrades by compression for any value of P due to change of code placement. Thus, code compression may improve the WCET and the energy consumption while reducing the code size, but fine-tuning of each application must be carried out to find good trade-offs if possible since not any value of P leads to the improvement of all concerned criteria. Hence, there is a real need for a compression strategy designed for applications that are subject to various constraints to find good trade-off and to avoid degrading one or more criteria, in particular to avoid impairing cache analysis and so increasing the WCET. The use of information from WCET analysis to fill the dictionary could favor the compression of instructions that have an important impact on the WCET which may improve the instruction cache analysis and so improve the WCET[14].

5.4 Impact of function inlining

Function inlining is a compiler transformation that replaces calls to functions with their bodies. This removes the call/return instructions as well as the prologue/epilogue code introduced in each function by the compiler. This is likely to improve performance. On the other hand, replicating function bodies increases the code size and degrades the temporal locality for accesses to the instruction cache, which might have a negative effect on the execution time. These expected positive/negative effects do not only stand for the average performance: they also concern the worst-case performance. However, as far as WCET estimation is concerned, removing control instructions, like calls and returns, is likely to improve the accuracy of results since, throughout the process of WCET analysis, control instructions are handled by join operations that introduce overestimation.

In the MORE project, we have developed a plugin to control the gcc –finline-functions optimization through the GCC-ICI interface [8]. With this plugin, it is possible to select the functions that should or not be inlined by the compiler.

Table 3 gives the impact on the WCET of function inlining. As explained above, it is difficult to forecast the impact of this optimization on the average and worst-case execution time due to opposite effects: reduction of flow control against degradation of temporal locality for instructions. Experimental results show that inlining eventually improves WCET estimates. Here, the gain is moderate but this is related to the small cache size considered in the paper. A larger cache would help in benefiting from inlining.

| benchmark | impact on WCET |
|---|---|
| adpcm | -1.5% |
| compress | -6.5% |
| helico | -7.4% |

**Table 3: Impact of function inlining on WCET**

Table 4 shows the increase in the code size due to function inlining. It can be observed that this increase is really significant. This suggests that strategies to tradeoff between the code size expansion and the WCET improvement should be set up.

| benchmark | impact on code size |
|---|---|
| adpcm | +45.5% |
| compress | +44.5% |
| helico | +95.9% |

**Table 4: Impact of function inlining on the code size**

Finally, Table 5 shows the impact of function inlining on energy consumption. This impact is very different from one benchmark to another. The gain in energy is almost zero for compress and is small for adpcm, while it is very important for helico. Indeed, for all benchmarks, inlining significantly decreases the number of accesses in the instruction cache and to a lesser extent to the data cache, which leads to important energy gains. However, for adpcm and compress, inlining increases the miss rates for both the instruction cache and the data cache, which lessens the gain in energy. For helico, on the contrary, both miss rates decrease with inlining, thus further improving energy.

| benchmark | impact on energy |
|---|---|
| adpcm | -5.2% |
| compress | -0.9% |
| helico | -73.9% |

**Table 5: Impact of function inlining on energy consumption**

## 6. Conclusion

Embedded systems are often subject to various constraints on code size, power requirements, execution time, etc. To meet these constraints, it may be necessary to transform the code: the code size can be reduced using code compression techniques, the energy consumption can be lowered with various strategies, among which specific data placement algorithms, the worst-case execution time can be improved by limiting the amount of jumps. However, experimenting several possible transformations to determine those that help in meeting the requirements is a costly and time-consuming process.

In this paper, we have introduced a framework that was developed within the French ANR MORE project with the goal of hosting various transformations and measurement or analysis tools to facilitate the optimization process. As illustrated with various examples, this framework provides the facilities that make it possible to support new transformations or analyses with limited efforts. Experimental results have assessed the usability of the framework.

Using the framework, it is possible to select the transformations that improve the target criterion, and it is also possible to evaluate their effects on other criteria. This is very important since many systems are not subject to a single constraint but instead to a combination of several constraints.

Our experimental results suggest that, in this case, it is necessary to set up appropriate strategies to combine several transformations while searching a tradeoff between the target criteria. This point is currently addressed in the second part of the MORE project. We are indeed developing an engine for iterative optimizations that controls the application of various transformations to determine the best combining as a function of the system constraints.